\documentclass{PoS}
\usepackage{amsmath,amssymb,amsfonts,dsfont,slashed}
\usepackage{graphicx}
\usepackage{booktabs,tabulary,cite}
\usepackage{placeins}
\usepackage[varg]{txfonts}

\newcommand{\mN}{m_N}

\newcommand{\diff}{\text{d}}

\newcommand{\MeV}{\,\text{MeV}}
\newcommand{\GeV}{\,\text{GeV}}
\newcommand{\beq}{\begin{equation}}
\newcommand{\eeq}{\end{equation}}

\newcommand{\xxx}{\mathbf{x}}
\newcommand{\mc}{m_\chi}
\newcommand{\F}{\mathcal{F}}

\title{Dark-matter--nucleus scattering in chiral effective field theory}

\ShortTitle{Dark-matter--nucleus scattering in chiral effective field theory}

\author{\speaker{Martin Hoferichter}\\      
Institute for Nuclear Theory, University of Washington, Seattle, WA 98195-1550, USA\\
E-mail: \email{mhofer@uw.edu}}

\author{Philipp Klos\\
Institut f\"ur Kernphysik, Technische Universit\"at Darmstadt, 64289 Darmstadt, Germany,\\
ExtreMe Matter Institute EMMI, 
GSI Helmholtzzentrum f\"ur Schwerionenforschung GmbH, 
64291 Darmstadt, Germany\\
E-mail: \email{pklos@theorie.ikp.physik.tu-darmstadt.de}}

\author{Javier Men\'endez\\
Center for Nuclear Study, The University of Tokyo, 113-0033 Tokyo, Japan\\
E-mail: \email{menendez@cns.s.u-tokyo.ac.jp}}

\author{Achim Schwenk\\
Institut f\"ur Kernphysik, Technische Universit\"at Darmstadt, 64289 Darmstadt, Germany,\\
ExtreMe Matter Institute EMMI, 
GSI Helmholtzzentrum f\"ur Schwerionenforschung GmbH, 
64291 Darmstadt, Germany,\\
Max-Planck-Institut f\"ur Kernphysik, Saupfercheckweg 1, 
69117 Heidelberg, Germany\\
E-mail: \email{schwenk@physik.tu-darmstadt.de}}

\abstract{Chiral effective field theory allows one to calculate the response of few-nucleon systems to external currents, both for currents that can be probed 
in the Standard Model and ones that only exist in Standard-Model extensions. In combination with state-of-the-art many-body methods,
the constraints from chiral symmetry can then be implemented in nuclear structure factors that describe the 
response of atomic nuclei in direct-detection searches for dark matter. We review the present status of this approach, including the role of 
coherently enhanced two-body currents, the discrimination of dark matter candidates based on the nuclear response functions, and limits on Higgs-portal dark matter.}

\FullConference{The 9th International Workshop on Chiral Dynamics\\
		17--21 September 2018\\
		Durham, NC, USA}

\begin{document}

\section{Introduction}

If dark matter is composed of weakly interacting massive particles (WIMPs), a possible strategy for discovery proceeds via direct-detection experiments, in which
the nuclear recoil produced by the scatter of WIMPs off atomic nuclei is sought, see~\cite{Akerib:2016vxi,Amaudruz:2017ekt,Cui:2017nnn,Agnese:2017njq,Agnes:2018ves,XMASS:2018bid,Aprile:2018dbl} 
for some of the most recent exclusion limits.
However, the interpretation of these limits in terms of the properties of the WIMP requires as critical input the local density and velocity distribution of the Galactic halo
as well as the response functions of the nuclear target. In the simplest case the information on the dark-matter candidate can be subsumed into a single-nucleon cross section,
in terms of which the exclusion limits are typically presented as a function of the WIMP mass.
For the nuclear physics input, one needs nucleon matrix elements of the quark/gluon-level operators describing the interaction of the WIMP with Standard-Model fields to define 
effective interaction operators in terms of hadronic degrees of freedom, and then in a second step the embedding of these operators within the nuclear states of interest. 
Chiral symmetry can play a role in both: for WIMP--nucleus scattering, with reduced mass of the order of $\mu_{\mathcal{N}\chi}\sim 100\GeV$ and a relative velocity of $v\sim 10^{-3}$ the typical momentum transfer 
$q\leq 2 \mu_{\mathcal{N}\chi} v\sim 200\MeV$ is right in the vicinity of the pion mass, so that the effects of the spontaneous breaking of the chiral symmetry of QCD 
need to be incorporated in the analysis. In practice, chiral symmetry constrains the nucleon matrix elements and few-nucleon operators that describe the coupling of the WIMP to nucleons, consistently with the description of nuclear forces. Here, we will concentrate on the impact on the WIMP--nucleon interactions.
We use chiral effective field theory (EFT)~\cite{Epelbaum:2008ga,Machleidt:2011zz,Hammer:2012id}, which 
allows one to study not only the currents that exist in the Standard Model (vector, axial-vector), but also treat non-standard currents (scalar, pseudoscalar, tensor, $\ldots$) 
on the same footing based on their chiral realizations. The consequences of chiral symmetry for WIMP--nucleus scattering, to varying degree, have been studied in~\cite{Prezeau:2003sv,Cirigliano:2012pq,Menendez:2012tm,Klos:2013rwa,Cirigliano:2013zta,Hoferichter:2015ipa,Hoferichter:2016nvd,Bishara:2016hek,Gazda:2016mrp,Korber:2017ery,Bishara:2017pfq,Hoferichter:2017olk,Andreoli:2018etf,Hoferichter:2018acd}, in the following, we will discuss some selected aspects.

\section{Chiral effective field theory and dark matter}

\begin{figure}[t]
\centering
\includegraphics[width=0.8\linewidth]{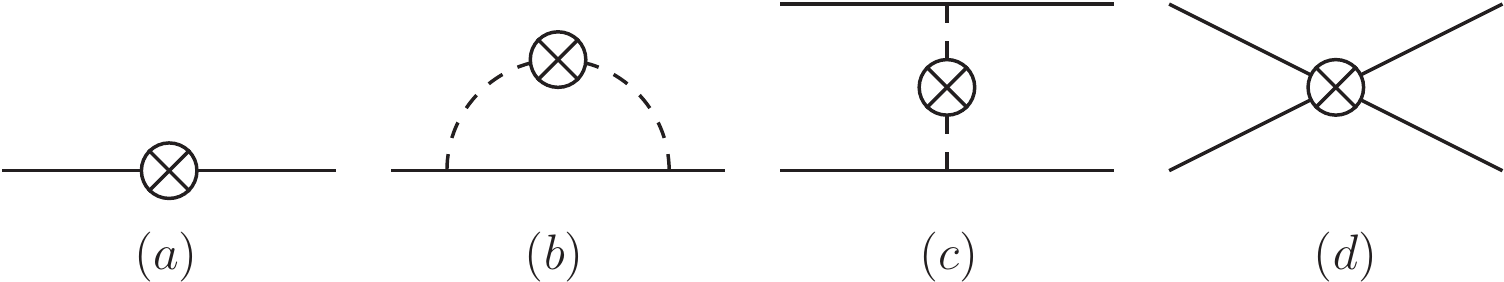}
\caption{Diagrams for the coupling of the WIMP current, indicated by the cross, to $(a)$ single nucleons, $(b)$ pion loops, $(c)$ pion-exchange diagrams, and $(d)$ two-nucleon contact terms.
}
\label{fig:diagrams}
\end{figure}

The analysis of WIMP--nucleus scattering in chiral EFT allows one to systematically derive the nuclear responses beyond the standard leading-order expressions---typically referred to as spin-independent (SI) and spin-dependent (SD) responses depending on whether the non-relativistic reduction gives a unity operator or the nucleon spin operator as in the $AA$ contribution 
in~\eqref{Lagr} below---while automatically implementing the constraints from the chiral symmetry of QCD. The different classes of corrections are shown in Fig.~\ref{fig:diagrams}: the single-nucleon contributions $(a)$ and $(b)$ essentially reproduce the known chiral expansion
of nucleon form factors, but the multi-nucleon diagrams $(c)$ and $(d)$ lead to genuinely new effects. Such two-body currents are well-established in electromagnetic and weak transitions of atomic nuclei~\cite{Gazit:2008ma,Menendez:2011qq,Bacca:2014tla,Pastore:2017uwc,Gysbers:2019uyb},
with currents that have been worked out up to one-loop order in the chiral expansion both for vector~\cite{Pastore:2008ui,Pastore:2009is,Kolling:2009iq,Pastore:2011ip,Kolling:2011mt,Krebs:2019aka}
 and axial-vector~\cite{Baroni:2015uza,Krebs:2016rqz} currents.
 
The starting point for the chiral analysis is then an effective Lagrangian for the interaction of the WIMP $\chi$ with quark and gluon fields, e.g., we give here
the quark operators for a spin-$1/2$ WIMP~\cite{Goodman:2010ku} 
\begin{align}
\label{Lagr}
 {\mathcal L}_{\chi}&=\frac{1}{\Lambda^3}\sum_q\Big[C_{q}^{SS}\bar \chi \chi \,m_q\bar q q
 +C_{q}^{PS}\bar \chi i\gamma_5 \chi \,m_q\bar q q 
 +C_{q}^{SP}\bar \chi \chi \,m_q\bar q i\gamma_5 q
 +C_{q}^{PP}\bar \chi i\gamma_5 \chi \,m_q\bar q i\gamma_5 q\Big]\notag\\
 &+\frac{1}{\Lambda^2}\sum_q\Big[C_q^{VV}\bar\chi\gamma^\mu\chi \,\bar q\gamma_\mu q
 +C_q^{AV}\bar\chi\gamma^\mu\gamma_5\chi\, \bar q\gamma_\mu q
 +C_q^{VA}\bar\chi\gamma^\mu\chi\, \bar q\gamma_\mu\gamma_5 q
 +C_q^{AA}\bar\chi\gamma^\mu\gamma_5\chi\, \bar q\gamma_\mu\gamma_5 q\Big],
\end{align}
where $\Lambda$ denotes the scale of physics beyond the Standard Model to make the Wilson coefficients $C$ dimensionless, the sums extend over the quark fields, and indices $S$, $P$, $V$, $A$ refer to the quantum numbers of the operators in the WIMP and quark bilinears. The chiral orders at which the various terms first arise are summarized in Table~\ref{table:counting}. The expected leading contributions
for SI and SD operators scale with $\nu=0$, $1$b $VV$ time-like and $AA$ space-like. The table reproduces the most important two-body corrections, in the $AA$~\cite{Menendez:2012tm,Klos:2013rwa} and 
$SS$~\cite{Prezeau:2003sv,Cirigliano:2012pq,Hoferichter:2016nvd,Hoferichter:2017olk,Hoferichter:2018acd} channels, but shows that relevant two-body currents can also arise for the $AV$ and $VA$ channels~\cite{Hoferichter:2015ipa}.

\begin{table}
\centering
\begin{tabular}{cccccccc}
\toprule
& Nucleon & & $V$ & & & $A$ &\\
WIMP & & $t$ & & $\xxx$ & $t$ & & $\xxx$ \\\midrule
& $1$b & $0$ & & $1+2$ & $2$ & & $0+2$\\
$V$ & $2$b & $4$ & & $2+2$ & $2$ & & $4+2$\\
& $2$b NLO & --- & & --- & $5$ & & $3+2$\\\midrule
& $1$b & $0+2$ & & $1$ & $2+2$ & & $0$\\
$A$ & $2$b & $4+2$ & & $2$ & $2+2$ & & $4$\\
& $2$b NLO & --- & & --- & $5+2$ & & $3$\\
\bottomrule
\end{tabular}
\ \ \
\begin{tabular}{cccc}
\toprule
& Nucleon & $S$ & $P$ \\
WIMP & & & \\\midrule
& $1$b & $2$ & $1$\\
$S$ & $2$b & $3$ & $5$\\
& $2$b NLO & ---& $4$ \\\midrule
& $1$b & $2+2$ & $1+2$\\
$P$ & $2$b & $3+2$ & $5+2$\\
& $2$b NLO & --- & $4+2$\\
\bottomrule
\end{tabular}
\caption{Chiral order $\nu$ of the one-body ($1$b), two-body ($2$b), and next-to-leading-order two-body ($2$b NLO) contributions in the various channels, separated into time-like and space-like components for the vector and axial-vector operators. The $+2$ indicates a suppression that arises from the non-relativistic expansion of the WIMP field if $\mc\sim\mN$, so that these terms can be ignored for a heavy WIMP. Table taken from~\cite{Hoferichter:2015ipa}.}
\label{table:counting}
\end{table}

The impact of chiral symmetry is particularly noteworthy in the scalar channel, where the first non-vanishing contribution only enters at $\nu=2$, the reason being that there is no scalar source in the leading-order 
pion--nucleon ($\pi N$) Lagrangian. For this reason, two-body currents are suppressed by only a single chiral order.
Moreover, chiral symmetry is a key ingredient in the phenomenological determination of the $1$b matrix element, the  $\pi N$ $\sigma$-term~\cite{Gasser:1990ce,Hoferichter:2015dsa,Hoferichter:2016ocj,RuizdeElvira:2017stg,Hoferichter_CD_piN}, via the Cheng--Dashen low-energy theorem~\cite{Cheng:1970mx,Brown:1971pn}. 
In particular, it is well-known that the strong $\pi\pi$ rescattering slows down the chiral expansion, in such a way that the full 
scalar form factor from dispersion theory should be employed instead~\cite{Gasser:1990ap,Hoferichter:2012wf}.

\begin{figure}[t]
\centering
\includegraphics[width=0.49\linewidth]{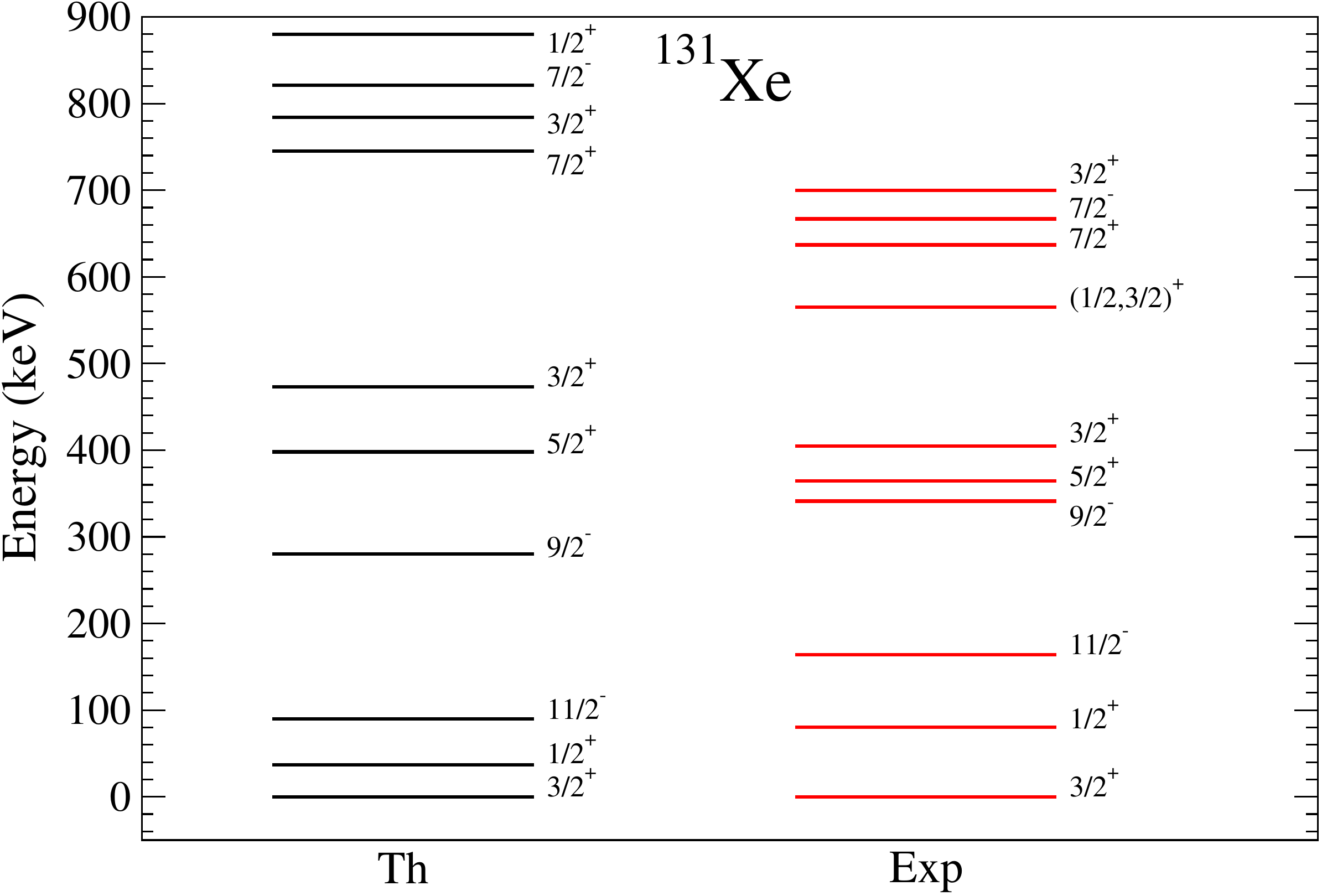}
\ \
\includegraphics[width=0.49\linewidth]{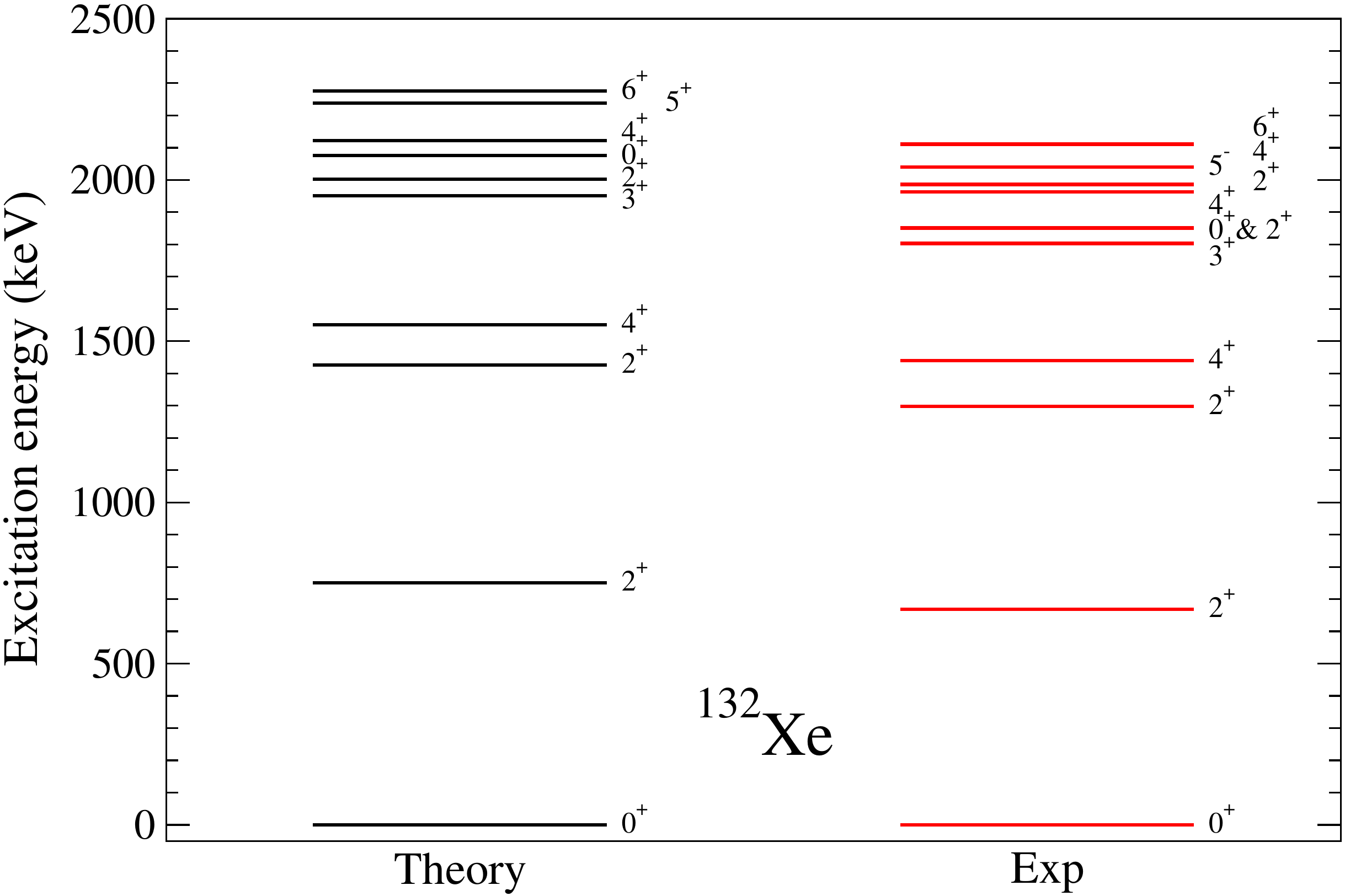}
\caption{Level spectra for $^{131}$Xe (left) and $^{132}$Xe (right). Figure taken from~\cite{Vietze:2014vsa}.
}
\label{fig:spectra}
\end{figure}

Ideally, the many-body methods used to calculate the nuclei of interest are based on consistent chiral interactions, but despite great progress in the study of electromagnetic and weak transitions in medium-mass nuclei~\cite{Hagen:2015yea,Parzuchowski:2017wcq,Morris:2017vxi,Gysbers:2019uyb}
such ab initio methods are not yet available for nuclei as heavy as xenon. Accordingly, the results presented here rely on the large-scale nuclear shell model~\cite{Caurier:2004gf}, which reproduces well nuclei relevant to direct-detection experiments, as shown for xenon in Fig.~\ref{fig:spectra}.
While currently we thus cannot rigorously quantify the nuclear-structure uncertainties, the diagnostics available in ab initio approaches should enable better uncertainty quantification in the future~\cite{Epelbaum:2014efa,Furnstahl:2015rha,Carlsson:2015vda}, again based on chiral EFT.

\section{Coherently enhanced two-body currents}

\begin{figure}[t]
\centering
\includegraphics[width=\linewidth,clip]{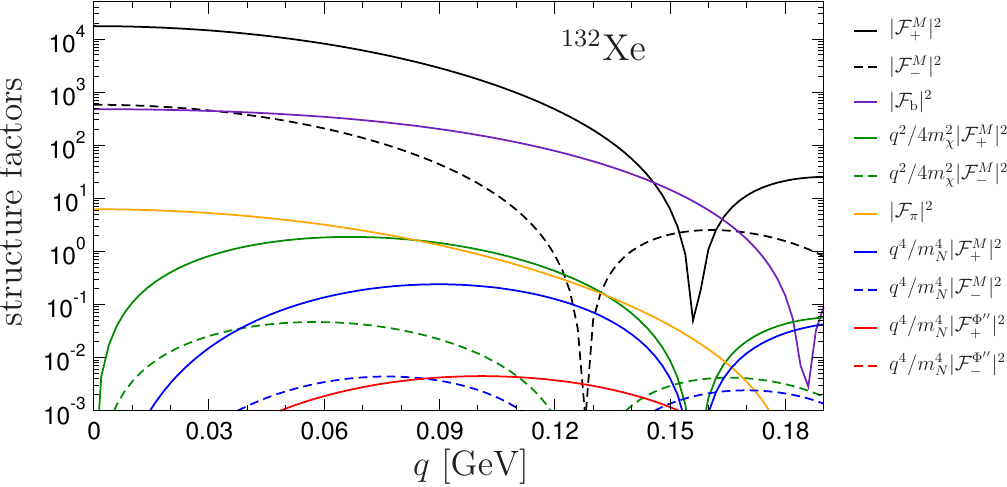}
\caption{Nuclear structure factors for $^{132}$Xe. Figure taken from~\cite{Hoferichter:2018acd}.
}
\label{fig:structure_factors}
\end{figure}

The scalar two-body currents are also special because they scale with the number of nucleons $A$ and thus inherit the coherent enhancement that characterizes the standard SI interaction,
and the same is true for a WIMP coupling with the trace anomaly $\theta^\mu_\mu$ or via a spin-$2$ operator.
In~\cite{Hoferichter:2016nvd,Hoferichter:2017olk,Hoferichter:2018acd} we developed the formalism to evaluate these two-body corrections in the nuclear shell model, including a consistent 
implementation of contact operators by renormalization to the nuclear binding energy.
In particular, this leads to a decomposition for the WIMP--nucleus cross section generalizing the standard SI expression~\cite{Hoferichter:2016nvd,Hoferichter:2018acd}
\begin{align}
\label{structure_factors}
\frac{\diff \sigma}{\diff q^2}&=\frac{1}{4\pi v^2}\bigg|\sum_{I=\pm}\Big(c_I^M-\frac{q^2}{m_N^2} \, \dot c_I^M\Big)\F_I^M(q^2)+c_\pi\F_\pi(q^2)
+c_\text{b}\F_\text{b}(q^2)+\frac{q^2}{2\mN^2}\sum_{I=\pm}c_I^{\Phi''}\F_I^{\Phi''}(q^2)\bigg|^2\notag\\
&+\frac{1}{4\pi v^2}\sum_{i=5,8,11}\bigg|\sum_{I=\pm}\xi_i(q,v^\perp_T)c_I^{M,i}\F_I^M(q^2)\bigg|^2,
\end{align}
where $q$ is the momentum transfer, $v$ the WIMP velocity, and the $c$ coefficients subsume nucleon matrix elements as well as the Wilson coefficients from~\eqref{Lagr}. Taking $^{132}$Xe as an example, our results for the 
nuclear structure factors $\F(q^2)$ are shown in Fig.~\ref{fig:structure_factors}. All our results are available as a \textsc{Python} package in the form of a \textsc{Jupyter} notebook, 
which can be downloaded from~\cite{notebook}.

\begin{figure}[t]
\centering
\includegraphics[width=0.7\linewidth,clip]{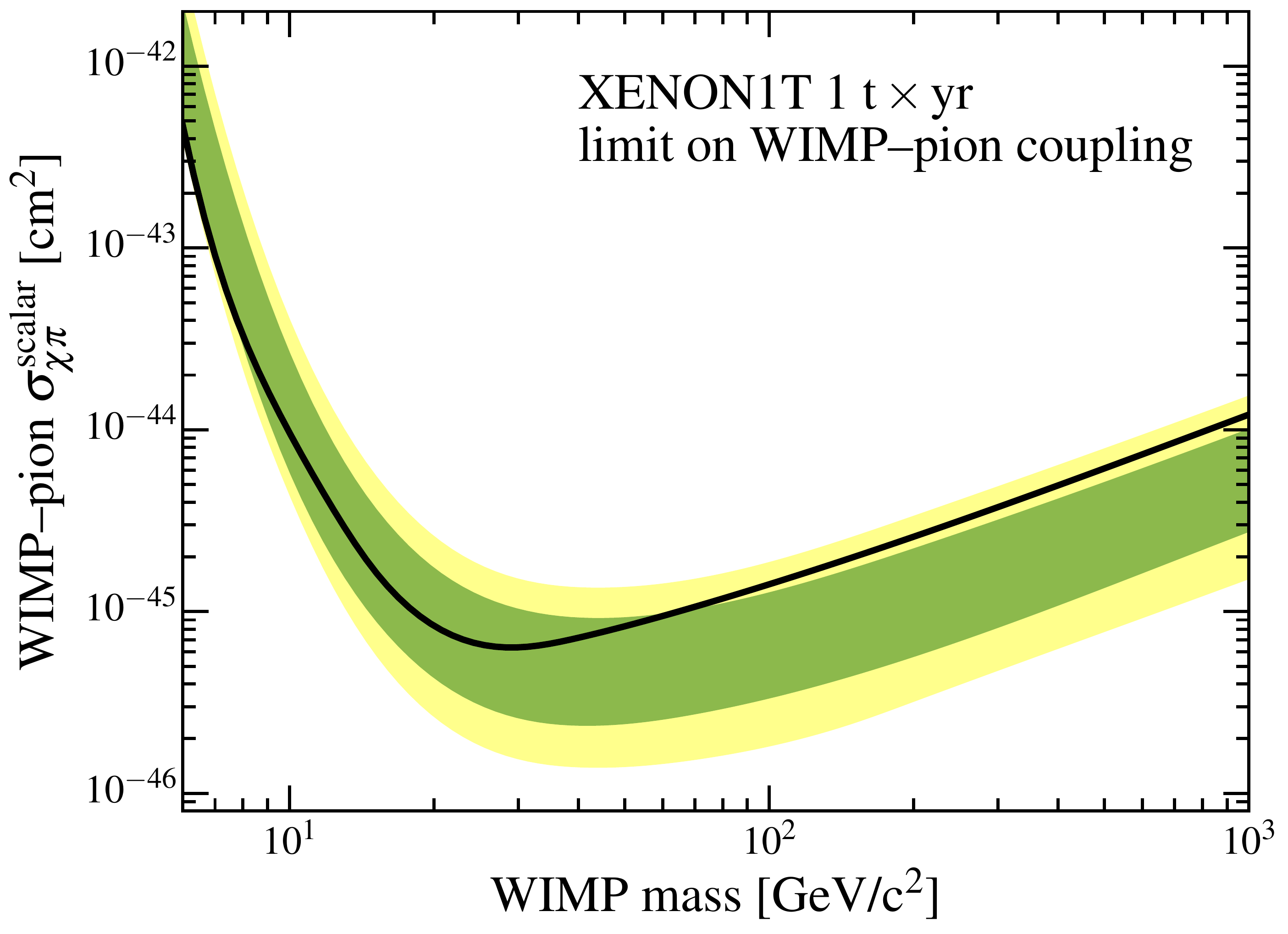}
\caption{Limits on the scalar WIMP--pion cross section from XENON1T. Figure taken from~\cite{Aprile:2018cxk}.
}
\label{fig:WIMPpion}
\end{figure}

The standard SI case is reproduced by omitting all terms in~\eqref{structure_factors} except for $c_+^M$: the relation 
$\sigma_{\chi N}^\text{SI}=\mu_N^2|c_+^M|^2/\pi$ establishes the connection to the usual formulation in terms of the single-nucleon cross section $\sigma_{\chi N}^\text{SI}$ (with reduced WIMP--nucleon mass $\mu_N$). Similarly, one can visualize some of the other terms by means of single-particle cross sections, most notably, if only $c_\pi$ is non-vanishing 
the WIMP only interacts with a virtual pion exchanged between the nucleons in the nucleus, see diagram $(c)$ in Fig.~\ref{fig:diagrams}.
Limits on $c_\pi$ derived under this assumption can then be interpreted in terms of a WIMP--pion cross section, providing a constraint in the space of 
all WIMP models complementary to the standard SI cross section. 
Such a constraint has recently been presented for the first time by the XENON collaboration~\cite{Aprile:2018cxk}, see Fig.~\ref{fig:WIMPpion}.
The argument is similar to searches for SD interactions~\cite{Fu:2016ega,Akerib:2017kat,Aprile:2019dbj,Amole:2019fdf}:
if the leading SI contribution is suppressed, e.g., in heavy-WIMP EFT~\cite{Hill:2013hoa} or in the context of so-called blind spots~\cite{Cheung:2012qy,Huang:2014xua,Crivellin:2015bva},
subleading corrections become more important and may dominate the nuclear response. Of these subleading responses the WIMP--pion coupling is coherently enhanced compared to the SD channel.

\section{Discriminating nuclear response functions}

\begin{figure}[t]
\centering
\includegraphics[width=0.7\linewidth,clip]{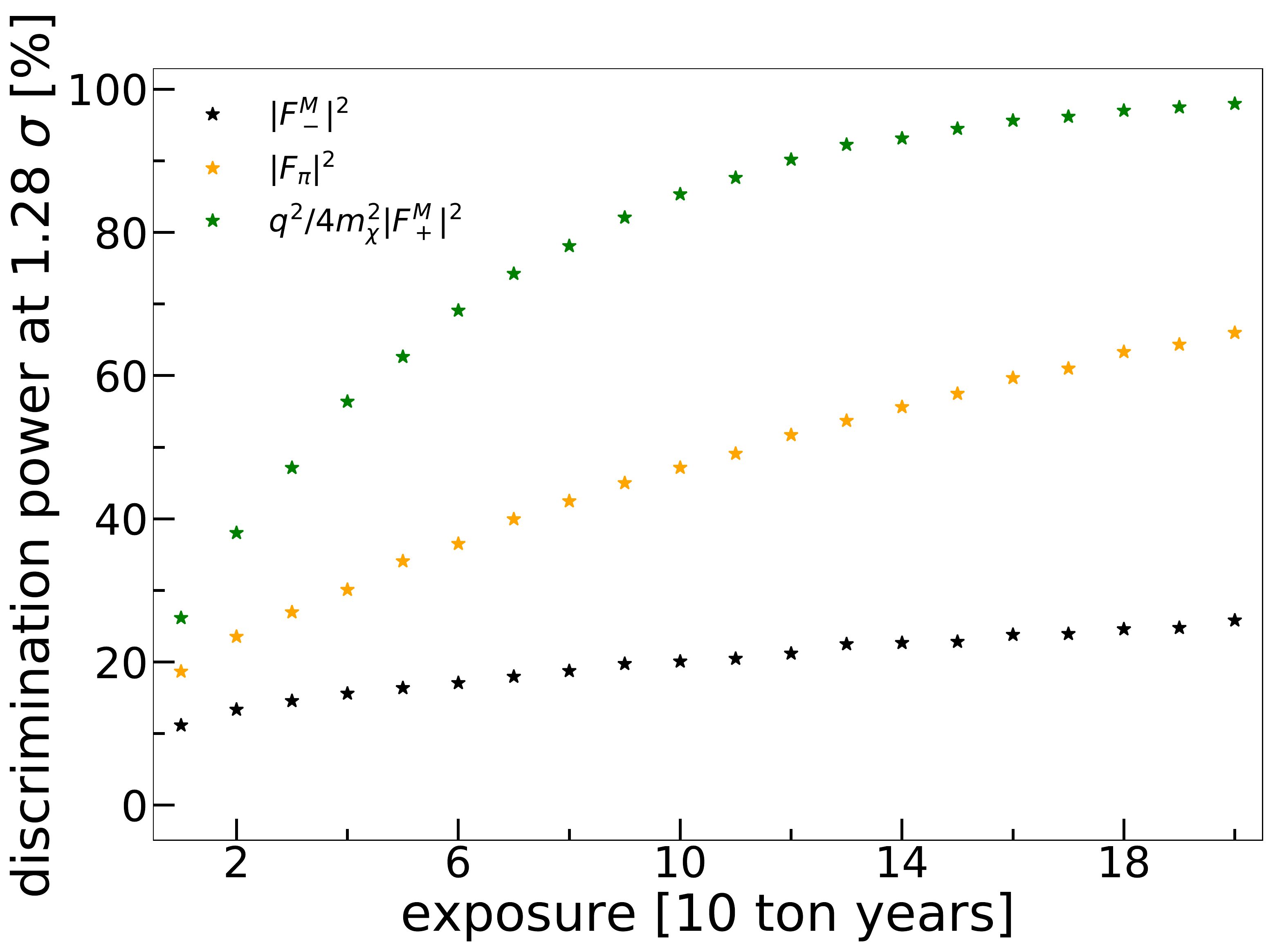}
\caption{Discrimination power with respect to $|\F_+^M|^2$ vs.\ exposure for three selected structure factors. The detector setting is DARWIN-like, with $\mc=100\GeV$ and interaction strength $\sigma_0=10^{-47}\,\text{cm}^2$. Figure taken from~\cite{Fieguth:2018vob}.
}
\label{fig:discrimination}
\end{figure}

Given the different terms in the decomposition~\eqref{structure_factors}, it is natural to ask how, in the case of a detection, the various contributions could be disentangled 
to gain insights into the nature of the WIMP. For instance, the separation of isoscalar and isovector terms could be achieved by comparing results for target materials with different $N/Z$ ratios.
Another strategy could rely on the $q$-dependence of the structure factors, which could be used to distinguish the different responses. 
In~\cite{Fieguth:2018vob} we studied to what extent such a discrimination would work in practice, starting from realistic detector settings for XENON100~\cite{Aprile:2012nq}, supplemented by
projections all the way to a potential DARWIN experiment~\cite{Aalbers:2016jon}.
A measure of the discrimination power with respect to the isoscalar SI response $\F_+^M$ vs.\ the exposure is shown in Fig.~\ref{fig:discrimination}, illustrating the key results from~\cite{Fieguth:2018vob}:
a separation via the $q$-dependence of the structure factors is possible in many cases as long as the overall interaction strength does not become too small. An exception concerns the
isovector SI response $\F_-^M$, which proves too similar to $\F_+^M$, see Fig.~\ref{fig:structure_factors}. In contrast, structure factors that vanish at $q=0$ are most easily differentiated,
but already for the scalar two-body currents encoded in $\F_\pi$ (relevant for the WIMP--pion coupling) the $q$-dependence does contain useful additional information.

\section{Limits on Higgs-portal dark matter}

\begin{figure}[t]
\centering
\includegraphics[width=0.9\linewidth,clip]{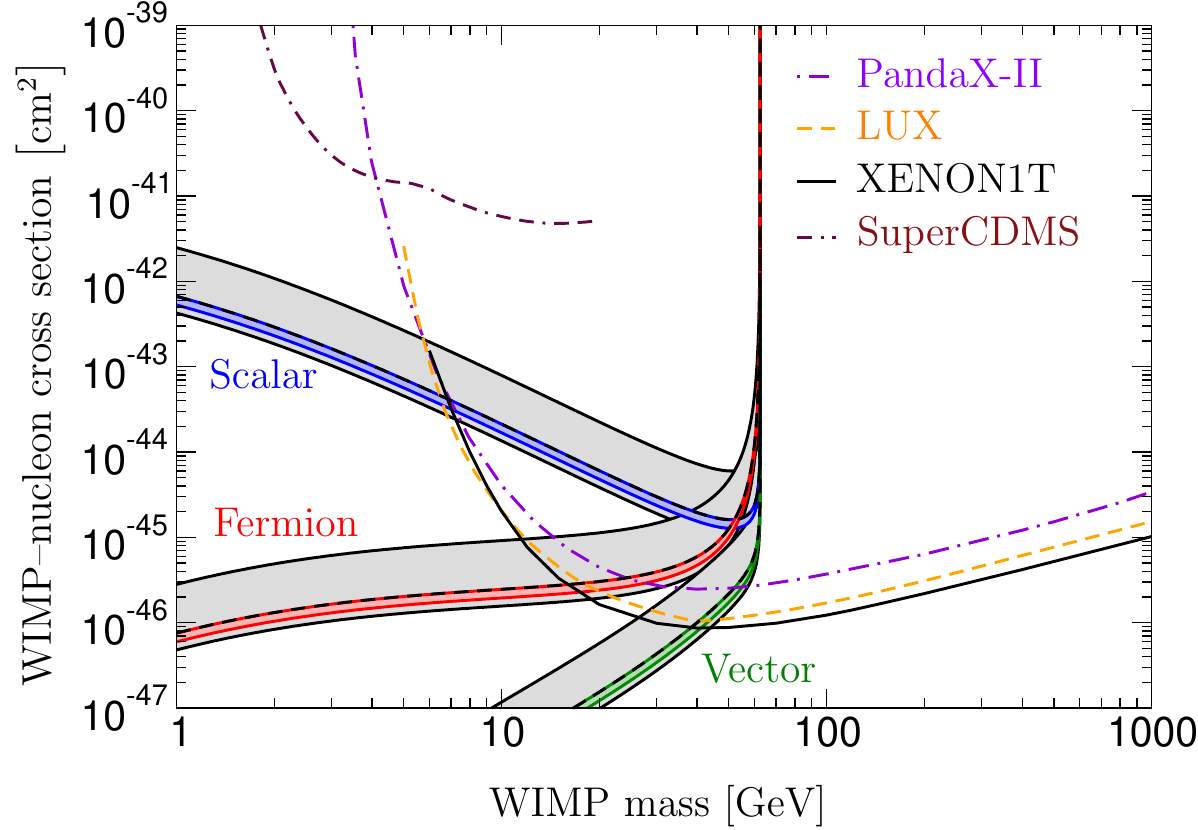}
\caption{Exclusion limits for scalar (blue), fermion (red), and vector (green) Higgs-portal WIMPs. The gray bands refer to the range $f_N=0.260\ldots 0.629$ from the ATLAS~\cite{Aad:2015pla} and CMS~\cite{Khachatryan:2016whc} analyses, the dashed lines to the central value $f_N=0.326$ considered therein, and the colored bands to our improved limits. 
 For comparison, we show the direct-detection limits from  SuperCDMS~\cite{Agnese:2015nto}, PandaX-II~\cite{Tan:2016zwf}, LUX~\cite{Akerib:2016vxi}, and XENON1T~\cite{Aprile:2017iyp}.
Figure taken from~\cite{Hoferichter:2017olk}.}
\label{fig:LHC}
\end{figure}

In Higgs-portal models for dark matter the WIMP interacts with the Standard Model via the exchange 
of the Higgs boson $H$. The WIMP itself can be either a scalar $S$, with interaction term ${\mathcal L}_S=H^\dagger H\, S^2$, a vector $V$, with ${\mathcal L}_V=H^\dagger H\, V_\mu V^\mu$, or a fermion $f$, 
with ${\mathcal L}_f=H^\dagger H\, \bar{f}f$, see, e.g.,~\cite{Kanemura:2010sh,Djouadi:2011aa,Djouadi:2012zc,Beniwal:2015sdl}.
If such couplings exist and if $\mc$ is less than half the mass of the Higgs boson, the Higgs should decay into a pair of dark-matter particles, and this process can be constrained by searching for so-called invisible Higgs decays~\cite{Aad:2015pla,Khachatryan:2016whc}. However, to be able to compare LHC and direct-detection limits, input for the 
Higgs coupling to the nucleon is required, for which typically the range $f_N=0.260\ldots 0.629$ had been employed, leading to the gray bands in Fig.~\ref{fig:LHC}. 

In~\cite{Hoferichter:2017olk} we pointed out that this input used for the scalar couplings was outdated, so that with modern input the translation of limits on invisible Higgs decays to direct-detection cross sections could be sharpened significantly. In addition, given that two-body currents are effectively subsumed in the single-nucleon cross sections in the analysis of direct-detection experiments, 
this effect needs to be included in the conversion of the LHC limits as well. Since the effects from the heavy quarks are effectively included in terms of the QCD trace anomaly $\theta^\mu_\mu$,
one needs both the scalar and the $\theta^\mu_\mu$ two-body currents. The latter receives contact-term contributions, see diagram $(d)$ in Fig.~\ref{fig:diagrams},
which we renormalize to the nuclear binding energy by making use of the fact that the contact terms are the same as the ones that appear in the leading nucleon--nucleon potential. 
Combining one- and two-body contributions, we obtain
\beq
f_N=f_N^\text{1b}+f_N^\text{2b}=0.308(18),
\eeq
leading to the colored bands in Fig.~\ref{fig:LHC}.

\section{Conclusions}

We have reviewed the application of chiral EFT to the interpretation of direct-detection searches for dark matter. 
In particular, chiral symmetry imposes constraints on 
both the nucleon matrix elements and the nuclear structure factors, both of which are required to cross nuclear and hadronic scales 
and eventually connect the measured limits on the WIMP--nucleus scattering rate to the properties of the WIMP. 
We presented several applications of the chiral-EFT formalism, ranging from limits on WIMP--pion interactions to improved limits on Higgs-portal dark matter.
The currently most comprehensive work on chiral-EFT-based structure factors~\cite{Hoferichter:2018acd,Klos:2013rwa} gives results for 
fluorine, silicon, argon, germanium, and xenon, including the matching relations for spin-$1/2$ and spin-$0$ WIMPs for a wide variety of effective operators. 
All results are also publicly available in a \textsc{Python} notebook~\cite{notebook}.

\section*{Acknowledgments}
This work was supported in part by the US DOE (Grant No.\ DE-FG02-00ER41132),
the ERC (Grant No.\ 307986 STRONGINT), the DFG through SFB 1245 (Projektnummer 279384907),
the Max-Planck Society, the Japanese Society for the Promotion of Science KAKENHI through grant 18K03639, 
MEXT as "Priority Issue on Post-K computer" (Elucidation of the fundamental laws and evolution of the universe), JICFuS,  
and the CNS-RIKEN joint project for large-scale nuclear structure calculations.

\end{document}